\documentclass[12pt,preprint]{emulateapj}
\usepackage{graphicx}

\shorttitle{The Habitable Zone Gallery}
\shortauthors{Stephen R. Kane \& Dawn M. Gelino}
\slugcomment{Submitted for publication in PASP}

\begin{document}

\title{The Habitable Zone Gallery}
\author{Stephen R. Kane, Dawn M. Gelino}
\affil{NASA Exoplanet Science Institute, Caltech, MS 100-22, 770
  South Wilson Avenue, Pasadena, CA 91125, USA}
\email{skane@ipac.caltech.edu}


\begin{abstract}

The Habitable Zone Gallery ({\tt www.hzgallery.org}) is a new service
to the exoplanet community which provides Habitable Zone (HZ)
information for each of the exoplanetary systems with known planetary
orbital parameters. The service includes a sortable table with
information on the percentage of orbital phase spent within the HZ,
planetary effective temperatures, and other basic planetary
properties. In addition to the table, we also plot the period and
eccentricity of the planets with respect to their time spent in the
HZ. The service includes a gallery of known systems which plot the
orbits and the location of the HZ with respect to those orbits. Also
provided are animations which aid in orbit visualization and provide
the changing effective temperature for those planets in eccentric
orbits. Here we describe the science motivation, the under-lying
calculations, and the structure of the web site.

\end{abstract}



\section{Introduction}
\label{introduction}

The field of exoplanets has undergone enormous diversification over
the past 20 years. The reasons for this include (but are not limited
to): new detection techniques, longer period baseline, smaller
mass/radii sensitivity, and atmosphere detection and modeling. As a
result of these developments, we are able to accurately characterize
the orbits of exoplanets and infer properties of their atmospheres
surface conditions. The sensitivity of radial velocity and transit
surveys to planets at longer periods is fundamentally limited by the
survey duration. Many of the radial velocity surveys are now pushing
this detection threshold beyond orbital periods of 10 years and with
upgraded instruments are sensitive to masses only a few times that of
the Earth. This means that many of the known planets pass through or
remain in the Habitable Zone (HZ) of their parent stars, some of these
with potentially rocky surfaces or terrestrial sized moons.

Keeping an accurate and exhaustive list of the known exoplanets is an
increasingly difficult tasks, but there exist numerous electronic
sources of information regarding exoplanets and their host stars.
Notable examples are the Extrasolar Planets Encyclopeadia\footnote{\tt
  http://exoplanet.eu/}, the NASA Exoplanet Archive\footnote{\tt
  http://exoplanetarchive.ipac.caltech.edu/}, and the Exoplanet Data
Explorer (EDE)\footnote{\tt http://exoplanets.org/}. The EDE stores
information only for those planets which have complete orbital
solutions, typically from a Keplerian fit to radial velocity data
acquired on those targets, and thus is very useful for this study.

The Habitable Zone Gallery (HZG) is a unique service which tracks the
orbits of exoplanets in relation to the HZ of their host stars. This
includes calculation of planetary temperatures, percentage of the
orbital phase spent within the HZ, and figures/movies which depict
planetary orbits with respect to the HZ. This allows for easy
reference of interesting systems in this context and target selection
for future investigations. This is also a useful tool for describing
exoplanets in Education \& Public Outreach (EPO) efforts. Here we
detail the methodology used to perform the necessary calculations for
generating the information on the HZG and the figures presented
throughout.


\section{Science Motivation}
\label{motivation}

The HZ is a key concept in our understanding of the conditions under
which basic life can form and survive. On our own planet we find
extreme conditions under which organisms are able to not only sustain
metabolic processes, but thrive and grow. Thus this understanding
informs our precepts on how life formed in our own Solar System and
also the possibility of similar processes in exoplanetary
systems. Although the concept of the HZ has been in the literature for
some time, it is only within the past 20 years that complex
atmospheric models have been developed to allow a rigorous examination
of its nature. In particular, the response of different atmospheres to
varying amounts of stellar flux allows the determination of HZ
boundaries for known exoplanetary systems. The pioneering work on this
was carried out by \citet{kas93} with subsequent papers producing more
analytical expressions for a variety of main sequence stars
\citep{jon10,kas93,sel07,und03}.

Exoplanet survey sensitivity is spreading in various directions; to
lower masses, to longer periods, and to later and earlier spectral
types. This broadening of the field opens up new areas to explore in
terms of the Habitable Zones of these configurations. The topic of the
HZ has become even more relevant in the wake of the results from the
Kepler mission. Approaches to investigating the HZ of Kepler stars has
been suggested by \citet{kal11} and many of the candidates released
are within the HZ of their host stars \citep{bor11a,bor11b}. The
announcements of the Kepler-20 system \citep{fre12,gau12} and
Kepler-22b \citep{bor11c} mark the crossing of a significant threshold
in narrowing down the search for an Earth mass/radius planet which
lies within the HZ of a solar-type star. Although these results inform
us greatly as to the frequency of habitable terrestrial planets, the
relative faintness of the Kepler sample makes more detailed follow-up
difficult.

A new area of habitability being explored is that of terrestrial moons
of giant exoplanets which lie within the HZ of their parent
stars. This has been discussed in the context of the Kepler mission
quite thoroughly \citep{kal10,kip09}. Based upon the statistics from
our own Solar System, it is natural to expect that many of the planets
discovered via the radial velocity method do indeed harbor moons of
various size and composition. For the giant planets themselves, great
progress is being made towards understanding the effective
temperatures of these bodies depending upon their albedos, thermal
response, and atmospheric circulation \citep{kan10,kan11}. Likewise
for the moons in these systems, the habitability prospects will depend
largely upon the conditions under which the primary planet is
subjected to.

The science motivation and use cases for the HZG can thus be
summarized as follows. 1. To provide an interactive method by which
users may visualize the orbits and Habitable Zones of known
exoplanetary orbits. 2. To provide tools, graphics, and movies which
can be easily imported into presentations to facilitate communication
of these concepts in both public and scientific contexts. 3. To
provide an interactive table tool which allows users to sort for
planets which spend substantial amounts of time within the HZ and thus
aid in target selection for further studies. 4. Ultimately, and of
particular interest to the authors, one would like to investigate the
habitability of exoplanets and exomoons whose total energy budget
varies with a cyclic nature, usually caused by orbital eccentricity
and consequently variable stellar flux.


\section{Habitable Zone and Orbits}
\label{hab}

In order for the HZG to function, it requires the ability to calculate
HZ parameters in a non-interactive fashion. Here we describe the
stellar and planetary information required and the subsequent
calculations performed.


\subsection{Necessary Information}

The HZG uses as input a variety of stellar and planetary parameters to
carry out the necessary calculations. The data for the planets listed
in the HZG are primarily extracted from the EDE, described in more
detail by \citet{wri11}. The full list of required parameters is shown
in Table \ref{inputtable}. As one might expect, not all required
parameters are available in all cases. In order to provide as complete
a list as possible, the HZG code attempts to fill in the data gaps
using published material on known stellar and planetary
properties. These are described in detail in Section
\ref{assumptions}.

\begin{deluxetable}{ll}
  \tablecaption{\label{inputtable} Required input parameters}
  \tablehead{
    \colhead{Parameter} &
    \colhead{Symbol}
  }
  \startdata
\multicolumn{2}{l}{Stellar} \\
\ \ Mass & $M_\star$ \\
\ \ Radius & $R_\star$ \\
\ \ Effective temperature & $T_\mathrm{eff}$ \\
\ \ Surface gravity & $\log g$ \\
\multicolumn{2}{l}{Planetary} \\
\ \ Mass & $M_p$ \\
\ \ Radius & $R_p$ \\
\ \ Period & $P$ \\
\ \ Semi-major axis & $a$ \\
\ \ Eccentricity & $e$ \\
\ \ Periastron argument & $\omega$ \\
  \enddata
\end{deluxetable}


\subsection{Assumptions and Missing Values}
\label{assumptions}

The planetary temperature calculations are based upon parameters which
aren't available in all cases, and thus assumptions are occasionally
made. The largest sources of uncertainties in the calculations are
caused by the generally unknown radii of both the host star and the
planet.  In cases where the radius of the star is not available from
other sources, we estimate the radius from the derived values of the
surface gravity $\log g$ using the relation
\begin{equation}
  \log g = \log \left( \frac{M_\star}{M_\odot} \right) - 2 \log
  \left( \frac{R_\star}{R_\odot} \right) + \log g_\odot
\end{equation}
where $\log g_\odot = 4.4374$ \citep{sma05}.

For non-transiting planets, a planetary radius measurement is
invariably not available. The mass-radius relationship of exoplanets
has been extensively studied, particularly with regards to composition
models such as that described by \citet{swi12}. To produce an estimate
of planetary radii, we fit a simple model to the available data for
confirmed transiting planets extracted using the EDE. The data are
current as of 3rd January 2012 and include 135 planets.

\begin{figure}
  \includegraphics[angle=270,width=8.2cm]{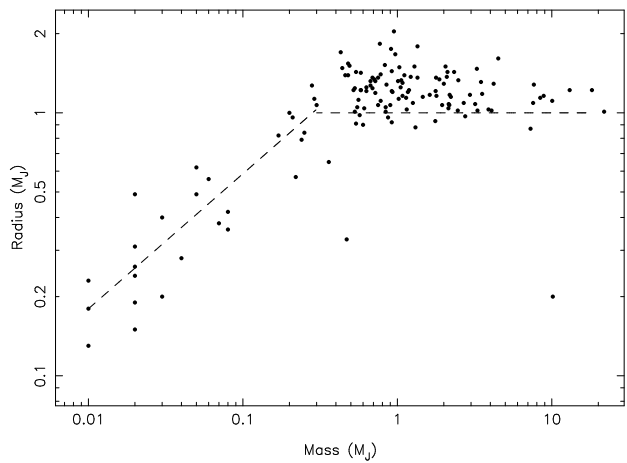}
  \caption{The mass and radius of known transiting planets. The dashed
    line indicates the model which the HZG uses to produce an
    approximation of the radius for non-transiting planets, where a
    power law is used for $M_p < 0.3 \ M_J$ and a constant value of
    $1.0 \ M_J$ is implemented for $M_p > 0.3$.}
  \label{massrad}
\end{figure}

These data are plotted in Figure \ref{massrad}. For masses greater
than $\sim 0.3$ Jupiter masses ($M_J$) the radii generally follow a
linear model which approximates to 1~$M_J$. This is true for masses
well into the brown dwarf regime (see for example for isochrone models
of \citet{bar03}). However, there is a significant divergence from a
linear model for masses less than 0.3~$M_J$. To account for this we
fit a power law to those data to obtain a better first order
approximation of the radii of those objects. The overall model used is
shown in Figure \ref{massrad} as a dashed line. These radii may be
used to calculate predicted planetary flux values at both infra-red
and optical wavelengths and also to estimate planetary densities for
aid in further characterization scenarios. This assumes that the
actual mass of the planets is close to the minimum mass deduced from
the Keplerian orbital solutions. It is also worth noting that this
sample is highly biased towards short-period orbits, typical of
transiting planets, where the planetary radii can be inflated due to
increased irradiation and tidal effects. Consequently, the median
period of the plotted sample is relative low (3.41~days).

A crucial parameter for estimating the planetary effective
temperatures is the planetary albedo, specifically at infrared
wavelengths. This is frequently assumed to be zero, as is the case
here, which then presupposes that the planet absorbs all incident
radiation and behaves as a blackbody. With the possible exception of
cases where internal and atmospheric heating contributes to this
temperature, the calculated temperature is likely to be
over-estimated.

Finally, many of the short-period planet discoveries (particularly
those discovered using the transit method) assume a circular orbit and
do not provide measurements of the eccentricity of argument of
periastron. In those cases we assume $e = 0.0$ and $\omega = 90\degr$
which is equivalent to fixing the time of periastron passage to the
predicted time of mid-transit.


\subsection{Fundamental Calculations}
\label{calculations}

For each of the planets which meet the minimum parameter requirements
described above, we perform the following calculations. The first step
is to calculate the extent of the HZ based upon the properties of the
host star. In order to do this we require the luminosity of the host
star, which is approximated as
\begin{equation}
  L_\star = 4 \pi R_\star^2 \sigma T_\mathrm{eff}^4
\end{equation}
where $\sigma$ is the Stefan-Boltzmann constant.

Using the boundary conditions of runaway greenhouse and maximum
greenhouse effects at the inner and outer edges of the HZ respectively
\citep{und03}, the stellar flux at these boundaries are given by
\begin{eqnarray*}
  S_\mathrm{inner} = 4.190 \times 10^{-8} T_\mathrm{eff}^2 - 2.139
  \times 10^{-4} T_\mathrm{eff} + 1.268 \\
  S_\mathrm{outer} = 6.190 \times 10^{-9} T_\mathrm{eff}^2 - 1.319
  \times 10^{-5} T_\mathrm{eff} + 0.2341
\end{eqnarray*}
The inner and outer edgers of the HZ are then derived from the
following
\begin{eqnarray*}
  r_\mathrm{inner} = \sqrt{ L_\star / S_\mathrm{inner} } \\
  r_\mathrm{outer} = \sqrt{ L_\star / S_\mathrm{outer} }
\end{eqnarray*}
where the radii are in units of AU and the stellar luminosities are in
solar units.

The next calculation performed is that of the planetary equilibrium
effective temperature. Without any direct knowledge of the surface
conditions or atmosphere of the planet, this calculation requires
numerous assumptions which can be used to estimate a range of
temperature values. One such assumption is that of the heat
redistribution efficiency of the atmosphere which depends upon zonal
wind speeds. If the atmosphere is 100\% efficient at redistributing
heat (``well-mixed'' model), the planetary equilibrium effective
temperature is given by
\begin{equation}
  T_p = \left( \frac{L_\star (1-A)}{16 \pi \sigma r^2}
  \right)^\frac{1}{4}
  \label{temp1}
\end{equation}
where $A$ is the spherical (Bond) albedo and $r$ is the star--planet
separation. If the atmosphere is completely inefficient with respect
to heat redistribution then the effective temperature is
\begin{equation}
  T_p = \left( \frac{L_\star (1-A)}{8 \pi \sigma r^2}
  \right)^\frac{1}{4}
  \label{temp2}
\end{equation}
which results in a hot dayside for the planet. The HZG provides the
calculated values for both the periastron and apastron locations of
the orbit. In the case of the movies (see Section \ref{contents}) the
well-mixed temperatures are calculated at each location of the orbit
where a Keplerian solution is evaluated.

The final calculation is that of the Keplerian orbit. The true
anomaly, $f$, is the angle between the position at periastron and the
current position in the orbit measured at the focus of the ellipse.
The mean anomaly is defined as
\begin{equation}
  M = \frac{2 \pi}{P} (t - t_p)
\end{equation}
and is hence the fraction of the orbital period that has elapsed since
the last passage at periastron, $t_p$. From the mean anomaly we can
calculate the eccentric anomaly, $E$, which is the angle between the
position at periastron and the current position in the orbit,
projected onto the ellipse's circumscribing circle perpendicularly to
the major axis, measured at the centre of the ellipse. These two
quantities are related via Kepler's equation
\begin{equation}
  M = E - e \sin E
\end{equation}
which we solve via the Newton-Raphson method and has the solution
\begin{equation}
  E = \frac{M - e (E \cos E - \sin E)}{1 - e \cos E}
\end{equation}

This yields the value of $E$ and hence the value of $f$
\begin{equation}
  \cos f = \frac{\cos E - e}{1 - e \cos E}
\end{equation}
The star--planet separation for eccentric planets has the following
form
\begin{equation}
  r = \frac{a (1 - e^2)}{1 + e \cos f}
\end{equation}
Thus, we deduce from Equations \ref{temp1} and \ref{temp2} that the
eccentricity of a planetary orbit introduces a time dependency to the
effective temperature of the planet.


\subsection{A Note on ``Habitability''}

The question arises as to whether those planets which are found to lie
within their host stars HZ are indeed habitable. This is unknown since
there are a variety of planetary properties that contribute towards
habitability in addition to the flux received from the host star
\citep{sch11}. For example, many of the planets described here are
giant planets which may not even have a rocky surface. Moons of those
systems may have conditions more suitable to sustaining life however
\citep{kan10}. The view of what makes a planet habitable has an
inevitable anthropic selection effect which contributes to our
selection of preferred targets \citep{wal11}. The survival of
extremophiles in orbits which are not consistently within the HZ are
also poorly understood since we have no precedents for those
conditions within our own system \citep{kan12}. The variety of factors
which thus influence the sustainability of life must be accounted for
when considering if a planet within the HZ is truely a candidate for
potentially hosting life.


\section{Web Site Structure}
\label{website}

Here we describe the overall structure and design of the web site and
how updates are carried out.


\subsection{Data Extraction and Sorting}
\label{extraction}

The HZG is currently not a stand-alone service but relies upon the
data curation of the EDE. The HZG code is executed periodically to
ensure that the catalog and data products remain in sync with the
services of the EDE. When executed, the code uses a 'wget' command to
extract all of the current data from the EDE and then identify the
columns of the output file which are needed by the HZG. These columns
are then sorted by the planet name and output to a new file ready to
be processed by the site construction code.

The HZG retains a copy of the previous data file for the purposes of
comparison. Specifically, the code checks for new planets or any
changes that may have occurred in the parameters for known planets. If
no changes are detected then the code halts at this point and any
alterations that may have been made are restored.


\subsection{Automated Reconstruction}

\begin{figure}
  \begin{center}
  \includegraphics[width=6.0cm]{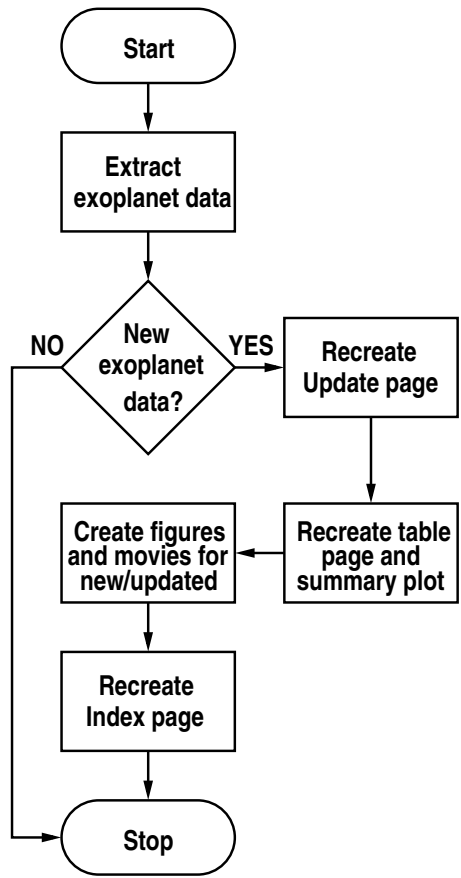}
  \caption{A flowchart which summarizes the automated reconstruction
    process wherever the HZG update code is executed.}
  \label{flowchart}
  \end{center}
\end{figure}

The philosophy of the HZG design is accuracy and sustainability such
that it can continue to provide useful results which keep pace with
the current rate of exoplanet detection. The optimal path to reach
this goal is to avoid data entry which dramatically increases
maintenance and the introduction of errors. The site update is
initiated via a single command which executes a shell script. As
described in Section \ref{extraction}, the script first retrieves the
most recent data and checks for changes. If there are new planets
and/or parameters then the script proceeds to execute a series of
FORTRAN programs which perform the necessary calculations and html
modifications to incorporate the additional information.

The flowchart shown in Figure \ref{flowchart} summarized how the
update and reconstruction of the web site functions. When changes to
the database are necessary, the update page is modified with new
planets added and/or changes to planetary parameters. The next step is
to update the summary plot shown on the main page and the table
page. It is computationally inexpensive to completely recreate these
items rather than attempt to only modify those values which have
changed. However, the next step of creating the figures in the gallery
and the movies is time-consuming and so this step is only initiated
where changes are required. The script finally recreated the index
(main) page before stopping. The script also has an optional flag
which, if used when executed, forces the complete reconstruction of
the entire site including all the figures and movies.

The revised version of the site is built at a temporary location where
the contents may be reviewed for accuracy and bug-related issues. Once
statisfied that the update has been successfully completed, the new
version is transferred to the operational location.


\subsection{The Table, Gallery, and Movies}
\label{contents}

The three main services provided by the HZG are a table of derived
planetary properties, a gallery of the planetary orbits with respect
the their HZ, and movies which provide a visual aid for the varying
predicted planetary temperatures.

The table includes the basic orbital properties of the planet, the
percentage of the orbital phase spent within the HZ, and calculated
temperatures at apastron and periastron for both the well-mixed and
hot-dayside models (see Section \ref{calculations}). The table is
sortable on any of these columns such that it can used to easily
select targets of interest. For example, shown in Table \ref{hzgtable}
is a subset of the complete table in the HZG which includes those
planets which spend 100\% of their time within their stars HZ. We have
only included the temperature columns in this truncated table. The HZG
additionally allows the user to click on any of the planet names to
load the location of that system with the Gallery section.

\begin{deluxetable}{lcccc}
  \tablecaption{\label{hzgtable} Some planets which spend 100\% of
    their time within the HZ.}
  \tablehead{
    \colhead{Planet} &
    \colhead{$T_\mathrm{eff}^a$ (K)} &
    \colhead{$T_\mathrm{eff}^b$ (K)} &
    \colhead{$T_\mathrm{eff}^c$ (K)} &
    \colhead{$T_\mathrm{eff}^d$ (K)}
  }
  \startdata
  tau Gru b   & 279 & 235 & 260 & 219 \\
  mu Ara b    & 322 & 271 & 283 & 238 \\
  Kepler-22 b & 340 & 286 & 340 & 286 \\
  HD 99109 b  & 296 & 249 & 271 & 228 \\
  HD 45364 c  & 341 & 286 & 309 & 260 \\
  HD 38801 b  & 349 & 294 & 349 & 294 \\
  HD 28185 b  & 326 & 274 & 310 & 261 \\
  HD 23079 b  & 302 & 254 & 272 & 229 \\
  HD 221287 b & 358 & 301 & 330 & 278 \\
  HD 188015 b & 328 & 276 & 286 & 240 \\
  HD 16760 b  & 288 & 243 & 270 & 227 \\
  HD 108874 b & 348 & 293 & 306 & 257 \\
  HD 10697 b  & 289 & 243 & 262 & 220 \\
  HD 10180 g  & 296 & 249 & 296 & 249 \\
  55 Cnc f    & 328 & 276 & 328 & 276
  \enddata
  \tablecomments{
$T_\mathrm{eff}^a$: periastron, hot-dayside;
$T_\mathrm{eff}^b$: periastron, well-mixed;
$T_\mathrm{eff}^c$: apastron, hot-dayside;
$T_\mathrm{eff}^d$: apastron, well-mixed.
}
\end{deluxetable}

\begin{figure*}
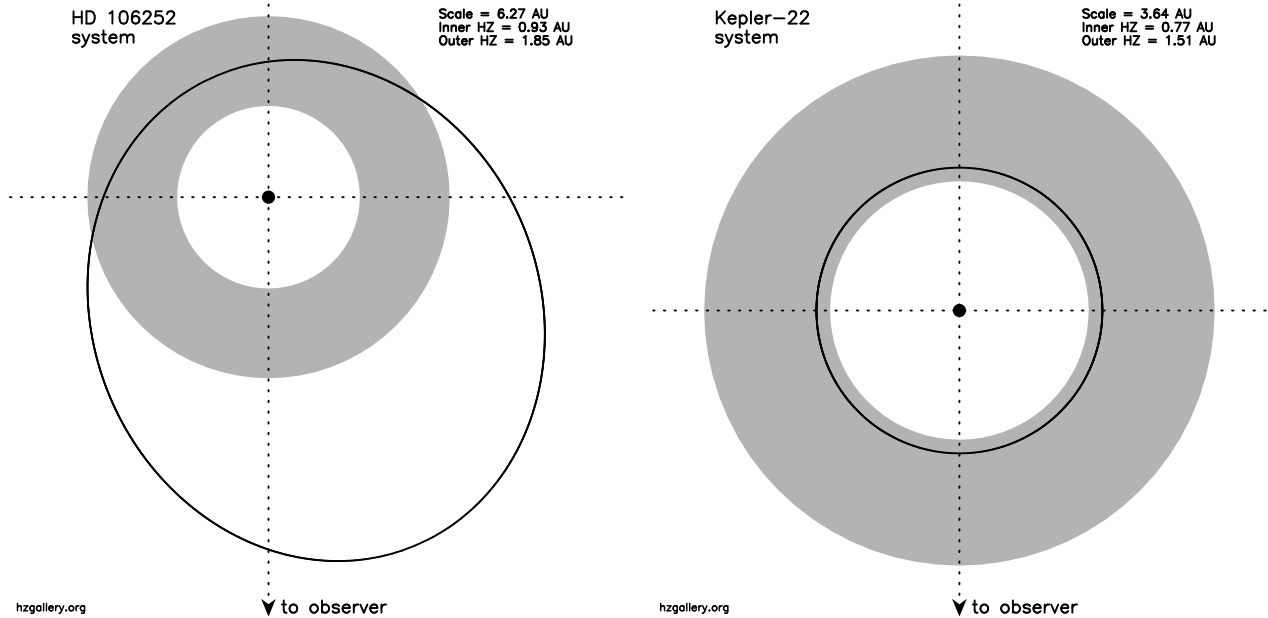

  \begin{center}
    \begin{tabular}{cc}
      \includegraphics[angle=270,width=8.2cm]{f03a.ps} &
      \includegraphics[angle=270,width=8.2cm]{f03b.ps}
    \end{tabular}
  \end{center}
  \caption{Two examples of orbits represented in the Gallery section
    of the HZG: HD~106252b (left) and Kepler-22b (right).}
  \label{galleryfig}
\end{figure*}

The Gallery section includes up to three plots per system depending in
the location of inner and outer planets with respect to their HZ. The
purpose of this is to accommodate the plotting of systems whose
planetary orbits occupy a wide range of orbital radii. In Figure
\ref{galleryfig} are shown two examples from the HZG which demonstrate
the diversity of these figures. HD~106252~b is in a 1531~day orbit
whose periastron occurs within the HZ of its star. Only 17\% of the
total orbital phase occupies the HZ though since the planet is moving
faster near periastron. This may be contrasted with the case of
Kepler-22~b, also shown in Figure \ref{galleryfig}. The discovery of
this planet was recently announced by \citet{bor11c}. Independent
calculations by the HZG code confirm that the planet does indeed spend
100\% of the orbit within the HZ.

Finally, the Movies section of the HZG provides users with animations
for individual planetary orbits, available in MPEG-2 and MPEG-4
formats. The animations track the star-planet separation and
calculated effective temperatures for the well-mixed atmospheric
model. This is particularly useful for eccentric orbits where the
temperature range can be substantial. The animations are divided into
200 frames of equal time increments per orbit which aids in the
animation resolution for highly eccentric orbits during periastron
passage.


\section{Conclusions}

The rate of exoplanet discoveries is continuing to increase, both with
the radial velocity and transit methods. As the sensitivity and time
baseline of the methods improve, the orbits of the discovered planets
are increasingly being found to lie in the HZ of their host stars or
beyond. This is a crucial step in identifying planets which are best
suited for follow-up activities related to habitability and
astrobiology.

The HZG seeks to provide a service in an easily maintainable way in
order to keep up with the discovery rate. The HZG mostly includes
planets whose host stars are relatively bright since they were
primarily discovered using the radial velocity technique. The
advantage of this is that these stars lend themselves to more
accessible follow-up observations which can be better used to
characterize those planets.

The structure of the site allows for automatic updates and
reconstruction in a very short period of time. The figures and movies
are intended for use in both scientific and public contexts and can be
easily transported into any presentation. The tools allow the user to
determine which known planets spend substantial time in their
Habitable Zones and to visualize their orbits and also perform general
investigations into the demographics of these targets. The HZG will
continue to adapt to the needs of the exoplanet community and further
develop tools as needed.


\section*{Acknowledgements}

The authors would like to thank Ravikumar Kopparapu, Lisa Kaltenegger,
and Jason Wright for their useful feedback on improvements to the
HZG. This research has made use of the Exoplanet Orbit Database and
the Exoplanet Data Explorer at exoplanets.org.



\begin{thebibliography}{}

\bibitem[\protect\citeauthoryear{Baraffe et al.}{2003}]{bar03}
  Baraffe, I., Chabrier, G., Barman, T.S., Allard, F., Hauschildt,
  P.H., 2003, A\&A, 402, 701
\bibitem[\protect\citeauthoryear{Borucki et al.}{2011a}]{bor11a}
  Borucki, W.J., et al., 2011, ApJ, 728, 117
\bibitem[\protect\citeauthoryear{Borucki et al.}{2011b}]{bor11b}
  Borucki, W.J., et al., 2011, ApJ, 736, 19
\bibitem[\protect\citeauthoryear{Borucki et al.}{2011c}]{bor11c}
  Borucki, W.J., et al., 2011, ApJ, in press (arXiv:1112.1640)
\bibitem[\protect\citeauthoryear{Fressin et al.}{2012}]{fre12}
  Fressin, F., et al., 2012, Nature, in press
\bibitem[\protect\citeauthoryear{Gautier et al.}{2012}]{gau12}
  Gautier, T.N., et al., 2012, ApJ, in press
\bibitem[\protect\citeauthoryear{Jones \& Sleep}{2010}]{jon10}
  Jones, B.W., Sleep, P.N., 2010, MNRAS, 407, 1259
\bibitem[\protect\citeauthoryear{Kaltenegger}{2010}]{kal10}
  Kaltenegger, L., 2010, ApJ, 712, L125
\bibitem[\protect\citeauthoryear{Kaltenegger \&
    Sasselov}{2011}]{kal11} Kaltenegger, L., Sasselov, D., 2011, ApJ,
  736, L25
\bibitem[\protect\citeauthoryear{Kane \& Gelino}{2010}]{kan10} Kane,
  S.R., Gelino, D.M., 2010, ApJ, 724, 818
\bibitem[\protect\citeauthoryear{Kane \& Gelino}{2011}]{kan11} Kane,
  S.R., Gelino, D.M., 2011, ApJ, 741, 52
\bibitem[\protect\citeauthoryear{Kane \& Gelino}{2012}]{kan12} Kane,
  S.R., Gelino, D.M., 2012, Astrobiology, submitted
\bibitem[\protect\citeauthoryear{Kasting et al.}{1993}]{kas93}
  Kasting, J.F., Whitmire, D.P., Reynolds, R.T., Icarus, 101, 108
\bibitem[\protect\citeauthoryear{Kipping et al.}{2009}]{kip09}
  Kipping, D.M., Fossey, S.J., Campanella, G., 2009, MNRAS, 400, 398
\bibitem[\protect\citeauthoryear{Schulze-Makuch et al.}{2011}]{sch11}
  Schulze-Makuch, D., 2011, Astrobiology, 11, 1041
\bibitem[\protect\citeauthoryear{Selsis et al.}{2007}]{sel07} Selsis,
  F., Kasting, J.F., Levrard, B., Paillet, J., Ribas, I., Delfosse,
  X., 2007, A\&A, 476, 1373
\bibitem[\protect\citeauthoryear{Smalley}{2005}]{sma05} Smalley,
  B., 2005, Mem. Soc. Astron. Ital. Suppl., 8, 130
\bibitem[\protect\citeauthoryear{Swift et al.}{2012}]{swi12} Swift,
  D.C., et al., 2012, ApJ, 744, 59
\bibitem[\protect\citeauthoryear{Underwood et al.}{2003}]{und03}
  Underwood, D.R., Jones, B.W., Sleep, P.N., 2003,
  Int. J. Astrobiology, 2, 289
\bibitem[\protect\citeauthoryear{Waltham}{2011}]{wal11} Waltham, D.,
  2011, Icarus, 215, 518
\bibitem[\protect\citeauthoryear{Wright et al.}{2011}]{wri11}
  Wright, J.T., et al., 2011, PASP, 123, 412

\end{thebibliography}
\end{document}